\documentclass[10pt, twocolumn]{IEEEtran}

\ifCLASSINFOpdf
  \usepackage[pdftex]{graphicx}
\else
\fi

\usepackage{amsfonts,amsmath,amsthm,amssymb,graphicx,epstopdf,cite,url,subfigure}
\usepackage{psfrag,amssymb,amsmath,pifont,cite,graphics,graphicx,epsfig,subfigure,amscd,helvet,multirow,enumerate}
\usepackage[pdftex]{graphicx}
\usepackage{color}
\usepackage{epstopdf}
\usepackage{accents}
\newtheorem{theorem}{Theorem}
\newtheorem{lemma}{Lemma}

\newtheorem{corollary}{Corollary}

\begin{document}

\title{A New Bound on Approximate Support Recovery}
\author{Hengkuan Lu and Jian Wang
\thanks{Lu and Wang are with the School of Data Science, Fudan University, Shanghai 200433, China (e-mail: \{hklu16, jian\_wang\}@fudan.edu.cn).}
}

\maketitle

\begin{abstract}
Orthogonal matching pursuit (OMP) is a greedy algorithm
popularly being used for the recovery of  sparse
signals. In this paper, we study the performance of OMP for support recovery of sparse signal under noise.  Our analysis shows that under mild constraint on the minimum-to-average ratio of nonzero entries in the sparse signal and the signal-to-noise ratio, the OMP algorithm can recover the support of signal with an error rate that can be arbitrarily small. Our result offers an affirmative answer to the conjecture of [Wang, TSP 2015] that the error rate of support recovery via OMP has no dependence on the maximum element of the signal.
\end{abstract}


{\IEEEpeerreviewmaketitle}

\maketitle
%
%

%
%

\section{Introduction} 
\IEEEPARstart{R}{e}cently, compressed sensing (CS) has attracted much attention in many fields, such as image processing, earth science and microwave imaging~\cite{donoho2006compressed,candes2005decoding}.
A fundamental problem in CS is to recover the support (i.e., identify the positions of non-zero elements) of a high-dimensional sparse signal from a small number of linear measurements
\begin{equation}
  \mathbf{y} = \mathbf{\Phi x} + \mathbf{v}
\end{equation}
where $\mathbf{x} \in \mathcal{R}^n$ is a $K$-sparse signal ($\|\mathbf{x}\|_0 \leq K \ll n$), $\mathbf{\Phi} \in \mathcal{R}^{m \times n}$ ($m < n$) is the measurement matrix, and $\mathbf{v}$ is noise. For years, much research has been devoted to seeking high-efficiency algorithms for support recovery. To avoid searching over all possible support for the optimal solution, which is NP-hard~\cite{natarajan1995sparse}, orthogonal matching pursuit (OMP) has gained significant popularity and become one of the most computationally efficient algorithms for performing this task~\cite{pati1993orthogonal,tropp2007signal}. In a nutshell, OMP identifies the support of signal in an iterative fashion, constructing a serial of support estimates that gradually fit the compressed measurements. Table \ref{tab:omp} offers a detailed description of OMP, where $supp(\mathbf{x}) = \{i:x_i\neq0\}$ denotes the support of $\mathbf{x}$, $\mathcal{T}^k$ signifies the estimated support in the $k$-th step, $\phi_i$ is the $i$th column of matrix $\mathbf{\Phi}$, and $\mathbf{r}^k$ is the residual vector in the $k$-th step.

\setlength{\arrayrulewidth}{1.5pt}
\begin{table} 
  \centering
      \caption[]{%
      The OMP Algorithm \label{tab:omp}}
      \vspace{-3mm}
\begin{tabular}{@{}ll}
\hline \\ \vspace{-14pt} \\
\textbf{Input}       &$\mathbf{\Phi}$, $\mathbf{y}$, and sparsity level $K$. \\
\textbf{Initialize}  & iteration counter $k = 0$, \\
                     & estimated support $\mathcal{T}^{0} = \emptyset$, \\
                     & and residual vector $\mathbf{r}^{0} = \mathbf{y}$.
                     \\
\textbf{While}       & $k < K$ \textbf{do}\\
                     & $k = k + 1$. \\
                     & Identify
                     \hspace{0.439mm}${t}^{k}  = \underset{i \in \Omega \backslash \mathcal{T}^{k - 1}}{\arg \max} |\langle \phi_i, \mathbf{r}^{k - 1} \rangle|$. \\
                     & Enlarge \hspace{0.58mm}$\mathcal{T}^{k} = \mathcal{T}^{k - 1} \cup t^{k}$. \\
                     & Estimate $\mathbf{x}^{k} = \underset{\mathbf{u}:\textit{supp}(\mathbf{u}) = \mathcal{T}^k}{\arg \min} \|\mathbf{y}-\mathbf{\Phi} \mathbf{u}\|_2$. \\
                     & Update \hspace{2.1mm}$\mathbf{r}^{k} = \mathbf{y} - \mathbf{\Phi} \mathbf{x}^{k}$. \\
\textbf{End}         \\
\textbf{Output}      & the estimated support $\mathcal{T}^{K}$ and vector $\mathbf{x}^K$.
\vspace{3pt} \\
\hline
\end{tabular}  
\vspace{-3mm}
\end{table}
\setlength{\arrayrulewidth}{1.3pt}

Both in theory and practice, it has been revealed that under appropriate conditions on the measurement matrix, the OMP algorithm produces reliable support recovery, while exhibiting high computational efficiency.  Specifically, it has been mathematically proved that perfect support recovery requires the signal-to-noise ratio (SNR) to approach infinity under the high dimensional setting, while fractional support recovery is not as hard~\cite{Galen2008}. For non-degenerate measurement matrices, upper and lower bounds on the probability of recovery error were given in \cite{Tang2009}, where the matrix $\mathbf{\Phi}$ is generated i.i.d at random and $\mathbf{v}$ is gaussian ($v_i \backsim \mathcal{N}(0,\sigma^2)$) . Unlike those probabilistic analyses,  an alternative direction of research was to build conditions for  exact support recovery~\cite{davenport2010analysis,liu2012orthogonal,livshits2012efficiency,wang2012Recovery,mo2012remarks,soussen2013joint}.
Among those, a popular framework for analysis is the restricted isometry property (RIP)~\cite{candes2005decoding}. A matrix $\mathbf{\Phi}$ is said to satisfy the RIP of order $K$ if there exists a constant $c \in [0, 1)$
such that
\begin{equation}
  \label{eq:RIP} (1- c) \| \mathbf{x} \|_2^2 \leq \| \mathbf{\Phi x} \|_2^2 \leq (1 + c) \| \mathbf{x} \|_2^2
\end{equation}
for all $K$-sparse vectors $\mathbf{x}$. In particular, the minimum value among all constants $c$ satisfying (\ref{eq:RIP}) is called the order-$K$ restricted isometry constant (RIC), noted as $\delta_{K}$. In the noise-free case (i.e., when the noise vector $\mathbf{v} = \mathbf{0}$), Davenport and Wakin showed that~\cite{davenport2010analysis} $\delta_{K + 1} < \frac{1}{3\sqrt K}$ is a sufficient condition for OMP to recover the whole support of the input signal accurately. Wang and Shim gave a shaper bound for OMP to perfectly recover a $K$-sparse signal: $\delta_{K + 1} < \frac{1}{\sqrt{K} +1}$~\cite{mo2012remarks,wang2012Recovery}, and after that it has been shown that
$\delta_{K + 1} < \frac{1}{\sqrt{K+1}}$ can also guarantee exact recover in $K$ iterations~\cite{wen2016}.


In many engineering applications of CS, noise can often be observed ($\mathbf{v} \neq \mathbf{0}$). In this senario, sufficient conditions for exact support recovery of input signals have been established~\cite{zhang2009consistency,wu2013exact,chang2014improved}, which depend on both the properties of the measurement matrices as well as the minimum magnitude of the non-zero elements of the signal. For $\ell_2$-bounded noise ($\| \mathbf{v} \|_2\leqslant\epsilon$ for some constant $\epsilon$) and $\ell_\infty$-bounded noise ($\|\mathbf{Av}\|_\infty\leqslant\epsilon$ for some constant $\epsilon$), the sharp bound for exact support recovery via OMP were given in \cite{wen2016}, in which the signal-to-noise ratio (SNR) is required to scale linearly with the sparsity level $K$. For high-dimensional settings, however, the SNR has to approach infinity, which is unrealistic for real applications.  

While exact support recovery requires an unbounded SNR, it has been shown that approximate support recovery (i.e., recovery of a fraction of support indices) can be guaranteed under a constant SNR~\cite{wang2015improved}. Consider the OMP algorithm running $K$ iterations before stopping and let $\rho_{\text{error}} := \frac{|\mathcal{T}^K \backslash \mathcal{T}|}{|\mathcal{T}|}$ denote the error rate of support recovery. Then if $snr \geq \kappa^2 \delta_{2K}^{-3/2}$, OMP recovers the support of $K$-sparse signal $\mathbf{x}$ from its noisy measurements $\mathbf{y}$ with error rate $\rho_{\text{error}} \leq C \kappa^2 \delta_{2K}^{1/2}$~\cite{wang2015improved},
where $\kappa := \max_{i,j \in  {supp}(\mathbf{x})} \frac{|x_{i}|}{|x_{j}|}$  and $C$ is a constant depending on $\delta_{2K}$. It has also been conjectured that the dependence on the maximum magnitude may be redundant, which arises from the intuition that the large element can maturely be picked up more easily. 
The purpose of this paper is to provide an affirmative answer to this conjecture. In particular,  our result is formally described in the following theorem.

\begin{theorem} \label{thm:1}
Let $mar:= \max_{i \in \text{supp}(\mathbf{x})} \frac{\|\mathbf{x}\|_{2}}{\sqrt{K} |x_{i}|}$ and $snr := \frac{\|\mathbf{\Phi x}\|_2^2}{\|\mathbf{v}\|_2^2}$. Then, for any constant $\rho_0\in(0,1)$, if $snr \geq \frac{C_{1} }{\rho_0 mar^2}$, where $C_{1}$ is a constant depending on $\delta_{2K}$, OMP recovers the support of $K$-sparse signal $\mathbf{x}$ from its noisy measurements $\mathbf{y} = \mathbf{\Phi x} + \mathbf{v}$ with error rate
$\rho_{\text{error}} \leq \rho_0$. 
\end{theorem} 

\section{Proof of Theorem~\ref{thm:1}} \label{sec:VI}
In this section, we give a detailed proof of Theorem~\ref{thm:1}.  
Before proceeding, we introduce some notations used in this paper. For $\Omega:= \{1,2, \cdots ,n\}$, let $\mathcal{T}:= \textit{supp}(\mathbf{x}) =\{i|i \in \Omega, x_{i} \neq 0\}$ denote the support of vector $\mathbf{x}$. For $\mathcal{S} \subseteq \Omega$,
$|\mathcal{S}|$ is the cardinality of set $\mathcal{S}$. $\mathcal{T}
\setminus \mathcal{S}$ is the set of all elements contained in
$\mathcal{T}$ but not in $\mathcal{S}$. For notational simplicity, let $\delta:= \delta_{2K}$. At the $k$-th iteration ($0 \leq k \leq K$) of OMP, let $\Gamma^k: = \mathcal{T} \backslash \mathcal{T}^k$ denote the set of missed detection of support indices. For given constant $\tau \in (0, 1]$, let $\Gamma_\tau^k$ denote the subset of $\Gamma^k$ corresponding to the $\lceil \tau K \rceil$ largest elements (in magnitude) of $\mathbf{x}_{\Gamma^k}$. Also, let $x_\tau^k$ denote the $\lceil \tau K \rceil$-th largest element (in magnitude) in $\mathbf{x}_{\Gamma^k}$. If $\lceil \tau K \rceil > |\Gamma^k|$, then set $\Gamma^k_\tau = \Gamma^k$ and $x_\tau^k = 0$. In this paper, we choose $\tau$ such that $\tau K$ is an integer (e.g., $\tau=1$ is a good choice).
 
Next, we provide some lemmas that are useful for our proof. 
 \begin{lemma}[Equations (35) and (60) in~\cite{wang2015improved}]\label{lem:1}
For any $0 \leq k \leq K - \lceil \delta^{1/2} K \rceil$, the residual of OMP satisfies
\begin{eqnarray}\label{eq:lem1}
 \|\mathbf{r}^k\|_2^2 - \|\mathbf{r}^{k + 1}\|_2^2  \geq   \frac{\|\mathbf{\Phi}' \mathbf{r}^k\|_\infty^2}{1 + \delta_{1}}  \geq   \frac{\langle \mathbf{r}^k ,\mathbf{\Phi}(\mathbf{w}-\mathbf{x}^k) \rangle^2 }{(1+\delta_1)\lceil\tau K\rceil\|\mathbf{x}_{\Gamma^K}\|^2_2}.
\end{eqnarray}
\end{lemma}

\begin{lemma}
  [Equation (61) in~\cite{wang2015improved}]\label{lem:2}
For any $0 \leq k \leq K - \lceil \delta^{1/2} K \rceil$, the residual of OMP satisfies
\begin{eqnarray} \label{eq:lem2}
&&\hspace{-7mm}\langle \mathbf{r}^k ,\mathbf{\Phi}(\mathbf{w}-\mathbf{x}^k) \rangle^2 \geq  \left ( (1-4\tau)\lceil \tau K\rceil(x^{k}_{\tau})^2- \Big( \frac{1}{\tau}-1 \Big) \|\mathbf{v}\|^2_2 \right) \nonumber\\ 
&&~~~~~~~~~~~~~~~~~~~~ \times\|\mathbf{\Phi}(\mathbf{w}-\mathbf{x}^k)\|^2_2.
\end{eqnarray}
\end{lemma}

We are now ready to present our main proof.  By the definition of SNR, we can rewrite \eqref{eq:lem2} as 
\begin{eqnarray}\label{eq:2}
&&\hspace{-6mm}\langle \mathbf{r}^k ,\mathbf{\Phi}(\mathbf{w}-\mathbf{x}^k) \rangle^2 \geq  \left ((1-4\tau)\lceil \tau K\rceil(x^{k}_{\tau})^2- \Big (\frac{1}{\tau}-1 \Big ) \frac{\|\mathbf{\Phi}\mathbf{x}\|^2_2}{snr} \right) \nonumber\\ 
&&~~~~~~~~~~~~~~~~~~~ \times\|\mathbf{\Phi}(\mathbf{w}-\mathbf{x}^k)\|^2_2.
\end{eqnarray}
Using  \eqref{eq:lem1} and \eqref{eq:2}, we have
\begin{eqnarray}\label{eq:4}
\lefteqn{\|\mathbf{r}^k\|_2^2 - \| \mathbf{r}^{k + 1}\|_2^2 \geq  \frac{1}{1+\delta_1}  \Big[(1-4\tau)\lceil \tau K\rceil(x^{k}_{\tau})^2 } \nonumber\\
  & & - \left (\frac{1}{\tau}-1 \right) \frac{\|\mathbf{\Phi}\mathbf{x}\|^2_2}{snr} \Big ]  \frac{\|\mathbf{\Phi}(\mathbf{w}-\mathbf{x}^k)\|^2_2}{\lceil \tau K\rceil\|\mathbf{x}_{\Gamma^K}\|^2_2}\nonumber\\
&\overset{(a)}\geq& \frac{1-\delta}{1+\delta_1} \Big [ (1-4\tau)(x^{k}_{\tau})^2 - \left (\frac{1}{\tau}-1 \right ) \frac{(1+\delta ) \|\mathbf{x}\|^2_2}{\lceil \tau K\rceil snr } 
\Big ]  \nonumber\\
&& \times \frac{\|(\mathbf{w}-\mathbf{x}^k)\|^2_2}{\|\mathbf{x}_{\Gamma^K}\|^2_2}\nonumber\\
&\overset{(b)} \geq& \frac{1-\delta}{1+\delta} \left [(1-4\tau)(x^{k}_{\tau})^2 - \Big (\frac{1}{\tau}-1 \Big ) \frac{(1+\delta)\|\mathbf{x}\|^2_2}{\lceil \tau K\rceil snr} \right],~~
\end{eqnarray}
where (a) is from the RIP, (b) is because $\delta :=\delta_{2K}\geq\delta_1$ (i.e.,  monotonicity of the RIP  constant~\cite{candes2005decoding}) 
and $$\|(\mathbf{w}-\mathbf{x}^k)\|^2_2\geq \|(\mathbf{w}-\mathbf{x}^k)_{\Omega\backslash\mathcal{T}^k}\|^2_2=\|\mathbf{x}_{\Gamma^K}\|^2_2.$$ Note that
\begin{eqnarray}
\lefteqn{\|\mathbf{r}^K\|_2^2  =  \|\mathbf{\Phi} (\mathbf{x} - \mathbf{x}^K) + \mathbf{v}\|_2^2} \nonumber \\  
& \overset{(a)}{\geq} & (1 - \theta_1) \|\mathbf{\Phi} (\mathbf{x} - \mathbf{x}^K)\|_2^2 - \left( {1}/{\theta_1} - 1 \right) \|\mathbf{v}\|_2^2 \nonumber \\
& \overset{(b)}{\geq} & (1 - \theta_1) (1 - \delta)\|\mathbf{x} - \mathbf{x}^K\|_2^2 - \left( {1}/{\theta_1} - 1 \right) \|\mathbf{v}\|_2^2 \nonumber \\
&{\geq} & (1 - \theta_1) (1 - \delta) \|(\mathbf{x} - \mathbf{x}^K)_{\Gamma^K}\|_2^2 - \left( {1}/{\theta_1} - 1 \right) \|\mathbf{v}\|_2^2 \nonumber \\
& \overset{(c)}{\geq} & (1 - \theta_1)(1-\delta) \|\mathbf{x}_{\Gamma^K}\|_2^2 - \left( {1}/{\theta_1} - 1 \right) \|\mathbf{v}\|_2^2. \label{eq:1}
\end{eqnarray} 
where (a) uses the fact that for any $\theta_1>0$, 
$$\|\mathbf{u} + \mathbf{v}\|_2^2 \geq (1 - \theta_1) \|\mathbf{u}\|_2^2 - \left( {1}/{\theta_1} - 1\right) \|\mathbf{v}\|_2^2$$
 with $\mathbf{u} = \mathbf{\Phi} (\mathbf{x} - \mathbf{x}^K)$. Here we would fix $\theta_1\in(0,1)$ , so that $1-\theta_1>0$, for further proof. And (b) follows from the RIP, and (c) is because $\mathbf{x}^K$ is supported on $\mathcal{T}^K$, and hence $\mathbf{x}^K_{\Gamma^K} = \mathbf{x}^K_{\mathcal{T} \backslash \mathcal{T}^K} = \mathbf{0}$. 
On the other hand,
\begin{eqnarray}
\|\mathbf{y}\|_2^2
& = & \|\mathbf{\Phi x + v}\|_2^2  \nonumber \\
& \overset{(a)}{\leq} & (1 + \theta_2) \|\mathbf{\Phi x}\|_2^2 + \left(1 + {1}/{\theta_2} \right) \|\mathbf{v}\|_2^2 \nonumber \\
& \overset{(b)}{\leq} & (1 + \theta_2) (1 + \delta) \|\mathbf{x}\|_2^2 + \left(1 + {1}/{\theta_2} \right) \|\mathbf{v}\|_2^2 , \label{eq:3}
\end{eqnarray}
where (a) is from the fact that for any $\theta_2>0$, 
$$\|\mathbf{u} + \mathbf{v}\|_2^2 \leq (1 + \theta_2) \|\mathbf{u}\|_2^2 +\left(1 + {1}/{\theta_2}\right) \|\mathbf{v}\|_2^2$$ with $\mathbf{u} = \mathbf{\Phi x}$, and (b) is due to the RIP.

Without loss of generality, assume that $\mathcal{T} = \{1, \cdots, K\}$ and that the elements of $\{x_i\}_{i = 1}^K$ are in a descending order of their magnitudes.  Then from the definition of $x_\tau^k$ we have that for any $k \geq 0$ and $k + \lceil \tau K \rceil \leq K$,\footnote{If $k + \lceil \tau K \rceil > K$, then $x_{k + \lceil \tau K \rceil}=0$.}
\begin{equation}
|x_\tau^k| \geq |x_{k + \lceil \tau K \rceil}|. \label{eq:22good}
\end{equation} 

Using \eqref{eq:3} and \eqref{eq:4}, we  have
\begin{eqnarray}\label{eq:5}
\|\mathbf{r}^K\|_2^2 
& = & \|\mathbf{r}^0\|_2^2 - \sum_{k = 0}^{K - 1} (\|\mathbf{r}^k\|_2^2 - \|\mathbf{r}^{k + 1}\|_2^2 ) \nonumber \\
&\leq& \|\mathbf{y}\|^2_2-\frac{(1-\delta)(1-4\tau)}{1+\delta}\sum_{k=0}^{K-1}(x^{k}_{\tau})^2\nonumber\\
&&+(\frac{1}{\tau}-1)\sum_{k=0}^{K-1}\frac{(1+\delta)\|\mathbf{x}\|^2_2}{\lceil \tau K\rceil snr }\nonumber\\
&\overset{(a)}\leq&(1 + \theta_2) (1 + \delta) \|\mathbf{x}\|_2^2 + \left(1 + {1}/{\theta_2} \right) \|\mathbf{v}\|_2^2 \nonumber\\
&&+ \frac{(1-\tau)(1+\delta)\|\mathbf{x}\|^2_2}{\tau^2 snr }-\frac{(1-\delta)(1-4\tau)}{1+\delta}\nonumber\\
&&\times\sum_{i=\lceil \tau K\rceil}^{K}(x_{i})^2,
\end{eqnarray} 
where (a) is because $\tau K$ is an integer and $$\sum_{k=0}^{K-1}(x^{k}_{\tau})^2 \geq  \hspace{-2mm} \sum_{k=0}^{K-\lceil \tau K\rceil}(x^{k}_{\tau})^2  \overset{\eqref{eq:22good}}{\geq}  \hspace{-.5mm}  \sum_{k=0}^{K-\lceil \tau K\rceil}(x_{k+\lceil \tau K\rceil})^2 = \hspace{-2mm} \sum_{i=\lceil \tau K\rceil}^{K}(x_{i})^2. $$ 

Furthermore, using \eqref{eq:1} and \eqref{eq:5}, we have
\begin{eqnarray}\label{eq:7}
 \lefteqn{ \left (1 - \theta_1 \right ) (1 - \delta) \|\mathbf{x}_{\Gamma^K}\|_2^2}\nonumber\\
&\overset{(a)}\leq& \left (\frac{1}{\theta_1}+\frac{1}{\theta_2} \right ) \frac{\|\mathbf{x}\|_2^2}{snr}+(1+\theta_2)(1+\delta)\|\mathbf{x}\|_2^2\nonumber\\
&& + \frac{(1-\tau)(1+\delta)\|\mathbf{x}\|^2_2}{\tau^2 snr}+\frac{4\tau(1-\delta)\|\mathbf{x}\|_2^2}{1+\delta}\nonumber\\
&&-\frac{1-\delta}{1+\delta}\sum_{i=\lceil \tau K\rceil}^{K} x_{i}^2\nonumber\\
&\overset{(b)}\leq& \left [\frac{1}{\theta_1}+\frac{1}{\theta_2}+\frac{(1-\tau)(1+\delta)}{\tau^2} \right] \frac{\|\mathbf{x}\|^2_2}{snr}\nonumber\\
&&+ \left [(1+\theta_1)(1+\delta)+\frac{4\tau(1-\delta)}{1+\delta} \right ] \|\mathbf{x}\|_2^2\nonumber\\
&&-K(1-\tau) \left ( \frac{1-\delta}{1+\delta} \right )  x_{min}^2.
\end{eqnarray}
where (a) is due to  that
\begin{eqnarray}\nonumber
\lefteqn{\frac{(1-\delta)(1-4\tau)}{1+\delta}\sum_{i=\lceil \tau K\rceil}^{K}x_{i}^2}\nonumber\\
 &=& -\frac{4\tau(1-\delta)}{1+\delta}\sum_{i=\lceil \tau K\rceil}^{K} x_{i}^2+\frac{1-\delta}{1+\delta}\sum_{i=\lceil \tau K\rceil}^{K} x_{i}^2\nonumber\\
&\geq& -\frac{4\tau(1-\delta)\|\mathbf{x}\|_2^2}{1+\delta}+\frac{1-\delta}{1+\delta}\sum_{i=\lceil \tau K\rceil}^{K} x_{i}^2 \nonumber
\end{eqnarray}
and (b) is because $\tau K$ is an integer.\\

Finally, by noting that $
\|\mathbf{x}_{\Gamma^K}\|_2^2 \geq |\Gamma^K| (x_{\min})^2$, and also  applying \eqref{eq:7}, we have
\begin{eqnarray}
\lefteqn{(1-\theta_1)(1-\delta)|\Gamma^K|  x_{\min}^2}\nonumber\\
 &\leq& (1-\theta_1)(1-\delta)\|\mathbf{x}_{\Gamma^K}\|_2^2\nonumber\\
&\leq& \left [ \frac{1}{\theta_1}+\frac{1}{\theta_2}+\frac{(1-\tau)(1+\delta)}{\tau^2} \right ] \frac{\|\mathbf{x}\|^2_2}{snr}+ \bigg [ (1+\theta_1)(1+\delta) \bigg. \nonumber\\
&& \bigg.  +\frac{4\tau(1-\delta)}{1+\delta} \bigg ] \|\mathbf{x}\|_2^2 - K(1-\tau)\frac{1-\delta}{1+\delta} x_{min}^2.
\end{eqnarray} 
That is, 
\begin{eqnarray}\label{eq:8}
 \frac{|\Gamma^K|}{K} &\leq&  \frac{1}{(1-\theta_1)(1-\delta)} \left [\frac{1}{\theta_1}+\frac{1}{\theta_2}+\frac{(1-\tau)(1+\delta)}{\tau^2} \right ]  \nonumber\\
&&\times \frac{\|\mathbf{x}\|^2_2}{snr K x_{min}^2}+\frac{1}{(1-\theta_1)(1-\delta)}\nonumber\\
&& \times \left [(1+\theta_1)(1+\delta)+\frac{4 \tau(1-\delta)}{1+\delta} \right ] \frac{\|\mathbf{x}\|_2^2}{K x_{min}^2}\nonumber\\
&&-\frac{1 - \tau}{(1-\theta_1)(1+\delta)}\nonumber\\
&=& C_1 \frac{\|\mathbf{x}\|^2_2}{snr K x_{min} ^2}+C_2\frac{\|\mathbf{x}\|_2^2}{K x_{min}^2}-C_3.
\end{eqnarray} 
Since $
\frac{C_3}{C_2}=\frac{1-\tau}{(1+\theta_1)(1+\delta)+\frac{4\tau(1-\delta)}{1+\delta}}\leq 1$ and 
$\frac{\|\mathbf{x}\|_2^2}{K(x_{min})^2}\geq 1$, we have 
$
C_2\frac{\|\mathbf{x}\|_2^2}{K(x_{min})^2}-C_3\geq 0 
$
and
\begin{eqnarray}
\rho_{\text{error}}=\frac{|\Gamma^K|}{K}\leq  C_1 \frac{\|\mathbf{x}\|^2_2}{snr  K x_{min}^2}=\frac{C_1}{snr \cdot mar}.\nonumber
\end{eqnarray}
Thus, for any constant $\rho_0\in(0,1)$, if  
$snr \geq \frac{C_1}{\rho_0  mar}$, 
then $\rho_{\text{error}}\leq \rho_0$. The proof is now complete.

\section{Discussions}
\subsection{Comparison to previous works}
\begin{figure}[t]
\centering
\includegraphics[scale=0.4]{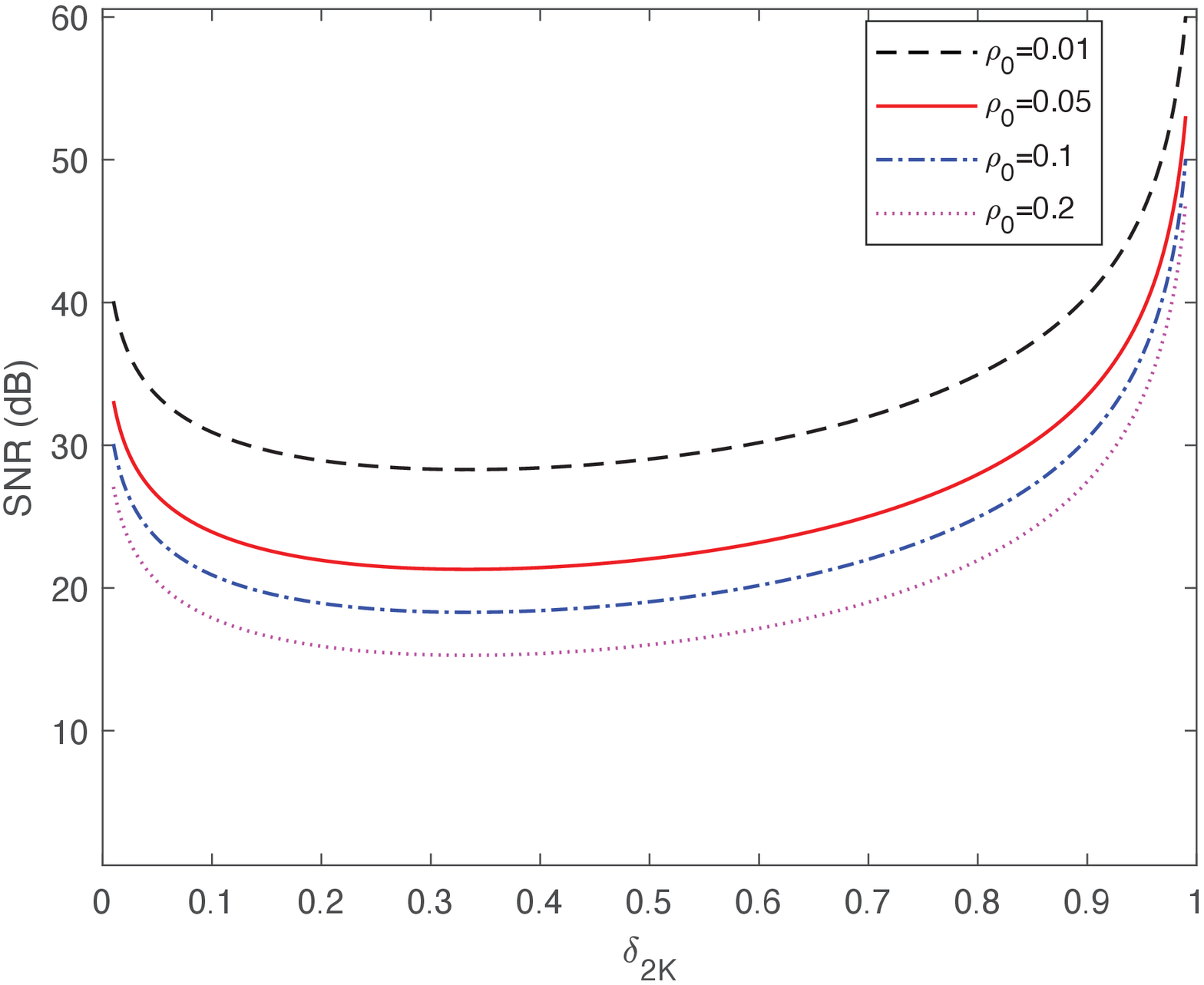}
\caption{The required $snr$ as a function of $\delta_{2K}$}
\label{fig:1}
\end{figure}

From \eqref{eq:8},  
\begin{equation}\nonumber
C_1 =\frac{1}{(1-\theta_1)(1-\delta)} \left [\frac{1}{\theta_1}+\frac{1}{\theta_2}+\frac{(1-\tau)(1+\delta)}{\tau^2} \right ].
\end{equation}
We fix $\theta_1=\delta$, $\tau=1$, and $\theta_2\rightarrow\infty$, so that $C_1 =\frac{1}{(1-\delta)^2}\frac{1}{\delta}$.
By fixing $\rho_0=\frac{\delta^{\frac{1}{2}}}{(1-\delta)^2}$, we have the following corollary. 
\begin{corollary} \label{corr:1}
Let $mar := \max_{i \in \text{supp}(\mathbf{x})} \frac{\|\mathbf{x}\|_{2}}{\sqrt{K} |x_{i}|}$. Then, if $snr \geq mar^2\delta^{-\frac{3}{2}}$, OMP recovers the support of $K$-sparse signal $\mathbf{x}$ from its noisy measurements $\mathbf{y} = \mathbf{\Phi x} + \mathbf{v}$ with error rate
\begin{equation}\label{cor:1}
\rho_{\text{error}} \leq \frac{\delta^{\frac{1}{2}}}{(1-\delta)^2}. 
\end{equation} 
\end{corollary} 
Let $\delta < 0.276$ so that $\rho_0\leq 1$. Then it can be shown that \eqref{cor:1} improves upon~\cite[Equation (7)]{wang2015improved}). First of all, the constant in~\eqref{cor:1} (i.e., $\frac{1}{(1-\delta)^2}$) is smaller than that in~\cite[Equation (43)]{wang2015improved} (i.e., $C=\frac{11}{1-2\delta^{\frac{1}{2}}}$. This is because, for any $\delta\in(0,1)$, $\frac{1}{(1-\delta)^2}<\frac{11}{1-2\delta^{\frac{1}{2}}}$ always holds. Second, $\max_{i \in  {supp}(\mathbf{x})} \frac{\|\mathbf{x}\|_{2}}{\sqrt{K} x_{i}|}$ is always smaller or equal to $\max_{i,j \in  {supp}(\mathbf{x})} \frac{|x_{i}|}{|x_{j}|}$. 
Moreover, consider signals with nonzero elements of equal magnitude (i.e., $mar =1$), we have the following result.  
\begin{corollary} \label{corr:2}
For any constant $\rho_0\in(0,1)$, if $snr \geq \frac{C_{1}}{\rho_0}$, where $C_{1}=\frac{1}{{\delta}(1-\delta)^2}$ is a constant, OMP recovers the support of $K$-sparse signal $\mathbf{x}$ from its noisy measurements $\mathbf{y} = \mathbf{\Phi x} + \mathbf{v}$ with error rate $\rho_{\text{error}} \leq \rho_0$.
\end{corollary} 
Fig.~\ref{fig:1} depicts the functional relationship between the required $snr$ and $\delta$ for different $\rho_0$.

Moreover, we would like to mention two major points of our proof that differ to those in \cite{wang2015improved}. 
\begin{enumerate}[i)]
\item  We fix $\tau$ so that $\tau K$ is an integer (e.g., $\tau=1$ simply meets the condition). By doing so, we can write $\lceil \tau K\rceil$ as $\tau K$, which can largely simplify the proof. In fact, we can fix $\tau$, $\theta_1$ and $\theta_2$ to any value in their feasible domain  to promote our proof. Since $0 < \theta_1 < 1$ and $\theta_2 > 0$, for any given $\delta$ and fixed $\tau$, we can fix $\theta_1$ and $\theta_2$ to make sure that  $C_1$ and is bounded. Whereas in \cite{wang2015improved}, $\delta$ was set to be $\delta^{\frac{1}{2}}$, and $\theta_1$ was fixed to $\theta$, which are clearly not optimal. 

\item Our results are given in terms of $mar$, while having no dependence on $x_{max}$ (i.e., the element of maximum magnitude in the signal $\mathbf{x}$). The key idea that allows to do so is that we replace all $\|\mathbf{v}\|_2^2$ with $\frac{\|\mathbf{\Phi x}\|_2^2}{snr}$ and  bound $\|\mathbf{\Phi x}\|_2^2$ with the RIP. While in the \cite{wang2015improved}, $\|\mathbf{v}\|_2^2$ is bounded by the inequality 
\begin{eqnarray}
\frac{2}{\tau}\|\mathbf{v}\|_2^2 \leqslant 2\tau^2(1+\tau^2)K(x_{min})^2,\nonumber
\end{eqnarray}
so that some $\|\mathbf{\Phi x}\|_2^2$ has to be upper bounded by $(1+\delta)K(x_{max})^2$, which inevitably causes loosening in the subsequent analysis.
\end{enumerate}

\begin{figure}[t]
\centering
\includegraphics[scale=0.4]{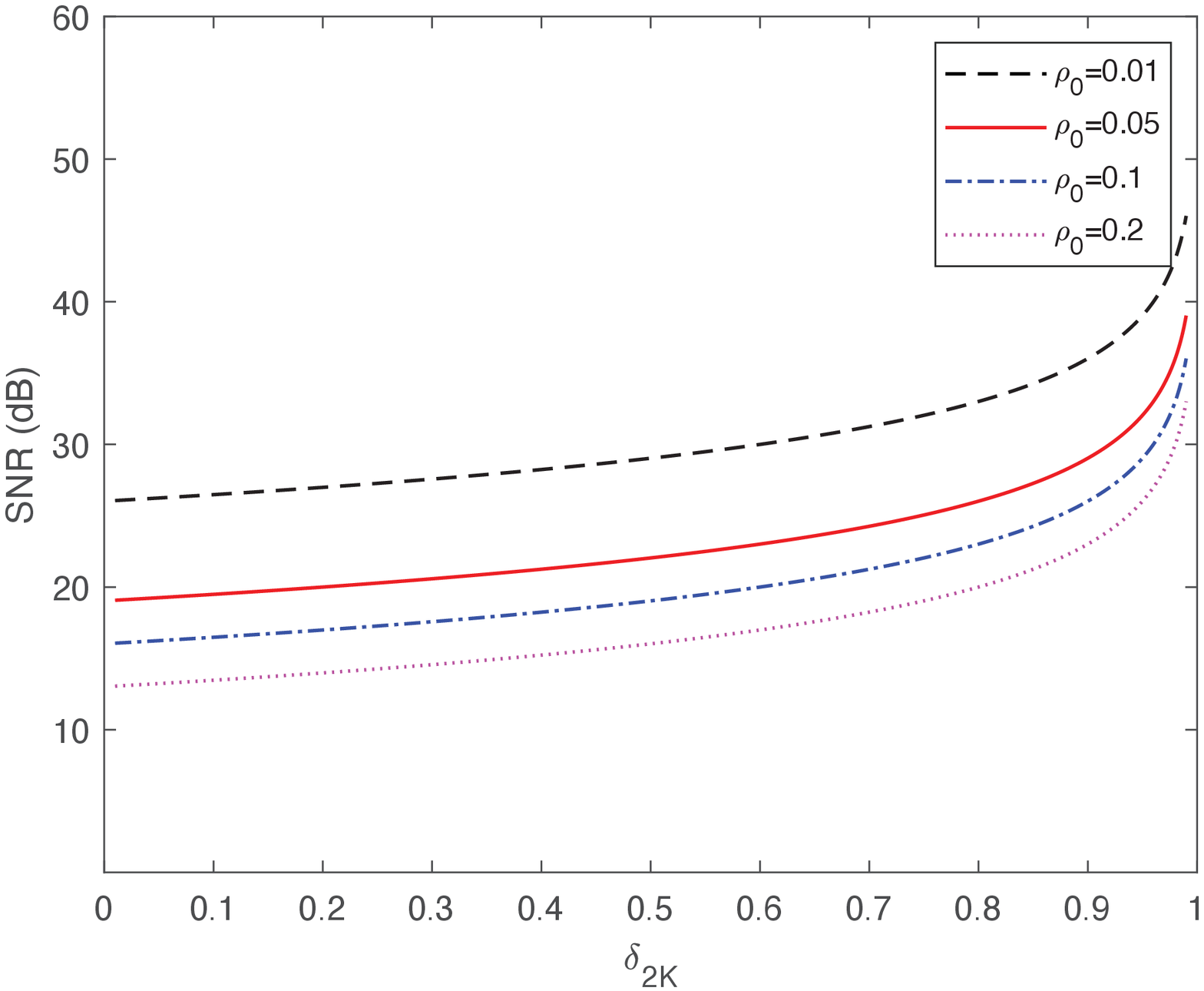}
\caption{The required $snr$ as a function of $\delta_{2K}$}
\label{fig:2}
\end{figure}   

\subsection{Best achievable bound}
To reach the best achievable result, we minimize $C_1$ in~\eqref{eq:8}. As $C_1 =\frac{1}{(1-\theta_1)(1-\delta)}[\frac{1}{\theta_1}+\frac{1}{\theta_2}+\frac{(1-\tau)(1+\delta)}{\tau^2}]$, fix $\theta_1=\frac{1}{2}$, $\tau=1$, $\theta_2\rightarrow\infty$ ($\frac{1}{\delta(1-\delta)}$ takes the minimum value when $\delta=\frac{1}{2}$) so that  $C_1 =\frac{4}{(1-\delta)}$. In the case that $mar =1$, we have the following corollary.  
\begin{corollary} \label{corr:3}
For any constant $\rho_0\in(0,1)$, if $snr \geq \frac{C_{1}}{\rho_0}$, where $C_{1}=\frac{4}{(1-\delta)}$ is a constant, OMP recovers the support of $K$-sparse signal $\mathbf{x}$ from its noisy measurements $\mathbf{y} = \mathbf{\Phi x} + \mathbf{v}$ with error rate $\rho_{\text{error}} \leq \rho_0$. 
\end{corollary} 

In Fig. \ref{fig:2}, we plot the functional relationship between the required $snr$ and $\delta$ for  different $\rho_0$. 

\subsection{Bounds on SNR and Sampling Rate}
Fix $C_1 =\frac{4}{(1-\delta)}$. Then, for any small error value $\rho_0$ we wish to achieve (i.e., $\rho_{\text{error}}=\rho_0$), it should satisfy that
\begin{eqnarray}\label{22}
snr\geq \frac{4}{\rho_0(1-\delta) mar}.
\end{eqnarray}
From~\cite[Equation (3)]{Galen2008} we have known that for any asymptotically reliable recovery must meet the conditions that:
\begin{eqnarray}\label{23_1}
\frac{m}{n}\geq\frac{h(\frac{K}{n})-h(\frac{K}{n},\rho_0)+\frac{1}{n}I(x;y|K)}{\frac{1}{2} {\log}(1+snr)}
\end{eqnarray}
where $h(x)=-x\textsl{log}(x)-(1-x)\textsl{log}(1-x)$ is the binary entropy function with
\begin{eqnarray}
h(x,y)=xh(y)+(1-x)h \left (\frac{y}{\frac{1}{x}-1} \right),
\end{eqnarray}
and $I(x;y|K)$ is mutual information between x and y conditioned on $K$, $\frac{m}{n}$ is the sampling rate. The conditional mutual information $I(x;y|K)$ is zero for non-stochastic signals, so the condition becomes
\begin{eqnarray}
snr\geq e^{\frac{2n}{m}(h(\frac{K}{n})-h(\frac{K}{n},\rho_0)}-1
\end{eqnarray}
To ensure an asymptotically reliable recovery,  $\rho_0$ should obey
\begin{eqnarray}\label{26}
\frac{4}{\rho_0(1-\delta) mar} \geq e^{\frac{2n}{m}(h(\frac{K}{n})-h(\frac{K}{n},\rho_0)}-1. 
\end{eqnarray}
Since $
\frac{4}{\rho_0(1-\delta)\cdot\text{MAR}}>\frac{4}{\rho_0},$
which is because $\delta \in (0,1)$ and $mar \leq 1$ and $
e^{\frac{2n}{m}(h(\frac{K}{n})-h(\frac{K}{n},\rho_0)}-1\leq e^{\frac{2n}{m}(h(\frac{K}{n})}-1$, 
(\ref{26}) can be guaranteed whenever
\begin{eqnarray}
\rho_0\leq \frac{4}{e^{\frac{2n}{m}h(\frac{K}{n})}-1}.
\end{eqnarray}
As $\rho_0<1$, this can be easily achieved by sparse signals for which $\frac{K}{n}$ is small enough.

On the other hand, from \cite[Equation (9)]{Galen2008}, we have known that an asymptotically reliable recovery is promised when the following inequalities hold:
\begin{equation} 
\begin{cases}
snr\cdot \rho_0 \geq e, \\
\frac{m}{n} > \frac{K}{n} + \frac{2h(\frac{K}{n})}{\textsl{log}(snr\cdot \rho_0 /e)}. 
\end{cases} \label{eq:jjjjffff}
\end{equation}
Since $\delta \in (0,1)$ and $mar \leq 1$, the first inequality in~\eqref{eq:jjjjffff} can be ensured by (\ref{22}). Also,  the second inequality of~\eqref{eq:jjjjffff} shows that if the sampling rate $\frac{m}{n}$ is fixed, it should be obeyed that 
\begin{eqnarray}\label{30}
snr\cdot \rho_0 > e^{\frac{2n\cdot h(\frac{K}{n})}{m-K}+1}. 
\end{eqnarray}
To get the second inequality of~\eqref{eq:jjjjffff} from (\ref{22}), we should have 
\begin{eqnarray}
\frac{4}{\rho_0(1-\delta) mar }\geq \frac{1}{\rho_0}e^{\frac{2n\cdot h(\frac{K}{n})}{m-K}+1}
\end{eqnarray}
Similarly, it can be guaranteed if  $
\frac{4}{\rho_0}\geq \frac{1}{\rho_0}e^{\frac{2n\cdot h(\frac{K}{n})}{m-K}+1},$
or equivalently, 
\begin{eqnarray}
e^{\frac{2n\cdot h(\frac{K}{n})}{m-K}+1}<4.
\end{eqnarray}
Again, this can be easily achieved by sparse signals with $\frac{K}{n}$ being small enough. Therefore, we can safely conclude that our results apply for asymptotically reliable recovery with achievable sampling rate.
%

\section{Conclusion}

In this paper, we have derived a new bound on the SNR for approximate support recovery via OMP. Our result improves upon that in \cite{wang2015improved}, while offering an affirmative answer to conjecture of whether one can remove the dependence of $x_{max}$ from the bound of SNR~\cite{wang2015improved}. In practice, our result indicates more stable evaluation about the SNR and error rate, especially when the signal of interest is mixed by impulse-like wave (e.g., a impulse wave mixed by some minute noise) whose $\max_{i,j \in \text{supp}(\mathbf{x})} \frac{|x_{i}|}{|x_{j}|}$ is unbounded from above. 

%


\begin{thebibliography}{10}

\bibitem{candes2005decoding}
E.~J. Cand{\`e}s and T.~Tao,
\newblock ``{Decoding by linear programming},''
\newblock {\em IEEE Trans. Inform. Theory}, vol. 51, no. 12, pp. 4203--4215,
  Dec. 2005.
  
  
\bibitem{donoho2006compressed}
D.~L. Donoho,
\newblock ``Compressed sensing,''
\newblock {\em IEEE Trans. Inform. Theory}, vol. 52, no. 4, pp. 1289--1306,
  Apr. 2006.




\bibitem{Galen2008}
G. Reeves and M. Gastpar,
\newblock ``{Sampling bounds for sparse support recovery in the presence of noise},''
\newblock{\em IEEE International Symposium on Information Theory}, pp. 2187-2191, July. 2008.


\bibitem{Tang2009}
G.  Tang and A. Nehorai, 
\newblock ``{Performance analysis for sparse support recovery},''
\newblock {\em IEEE Trans. Inform. Theory}, vol. 56, no. 3, pp. 1383-1399, 2010.



\bibitem{natarajan1995sparse}
B.~K. Natarajan,
\newblock ``Sparse approximate solutions to linear systems,''
\newblock {\em SIAM journal on computing}, vol. 24, no. 2, pp. 227--234, Apr.
  1995.



\bibitem{pati1993orthogonal}
Y.~C. Pati, R.~Rezaiifar, and P.~S. Krishnaprasad,
\newblock ``Orthogonal matching pursuit: Recursive function approximation with
  applications to wavelet decomposition,''
\newblock in {\em Proc. 27th Annu. Asilomar Conf. Signals, Systems, and
  Computers}. IEEE, Nov. Pacific Grove, CA, Nov. 1993, vol.~1, pp. 40--44.


\bibitem{zhang2009consistency}
T.~Zhang,
\newblock ``On the consistency of feature selection using greedy least squares
  regression,''
\newblock {\em J. of Mach. Learn. Res.}, vol. 10, pp. 555--568, 2009.



\bibitem{tropp2007signal}
J.~A. Tropp and A.~C. Gilbert,
\newblock ``{Signal recovery from random measurements via orthogonal matching
  pursuit},''
\newblock {\em IEEE Trans. Inform. Theory}, vol. 53, no. 12, pp. 4655--4666,
  Dec. 2007.





\bibitem{davenport2010analysis}
M.~A. Davenport and M.~B. Wakin,
\newblock ``{Analysis of Orthogonal Matching Pursuit using the restricted
  isometry property},''
\newblock {\em IEEE Trans. Inform. Theory}, vol. 56, no. 9, pp. 4395--4401,
  Sep. 2010.
  


%
%
  
  
\bibitem{liu2012orthogonal}
E.~Liu and V.~N. Temlyakov,
\newblock ``The orthogonal super greedy algorithm and applications in
  compressed sensing,''
\newblock {\em IEEE Trans. Inform. Theory}, vol. 58, no. 4, pp. 2040--2047,
  Apr. 2012.



\bibitem{livshits2012efficiency}
E.~D. Livshits,
\newblock ``On the efficiency of the orthogonal matching pursuit in compressed
  sensing,''
\newblock {\em Sbornik: Mathematics}, vol. 203, no. 2, pp. 183, 2012.



\bibitem{mo2012remarks}
Q.~Mo and Y.~Shen,
\newblock ``A remark on the restricted isometry property in orthogonal matching
  pursuit algorithm,''
\newblock {\em IEEE Trans. Inform. Theory}, vol. 58, no. 6, pp. 3654--3656,
  Jun. 2012.

\bibitem{wang2012Recovery}
J.~Wang and B.~Shim,
\newblock ``On the recovery limit of sparse signals using orthogonal matching
  pursuit,''
\newblock {\em IEEE Trans. Signal Process.}, vol. 60, no. 9, pp. 4973--4976,
  Sep. 2012.
  
\bibitem{soussen2013joint}
C.~Soussen, R.~Gribonval, J.~Idier, and C.~Herzet,
\newblock ``Joint $k$-step analysis of orthogonal matching pursuit and
  orthogonal least squares,''
\newblock {\em IEEE Trans. Inform. Theory}, vol. 59, no. 5, pp. 3158--3174, May
  2013.








\bibitem{chang2014improved}
L.~Chang and J.~Wu,
\newblock ``An improved \text{RIP}-based performance guarantee for sparse
  signal recovery via orthogonal matching pursuit,''
\newblock {\em IEEE Trans. Inform. Theory}, vol. 60, no. 9, pp. 5702--5715,
  Sep. 2014.

\bibitem{wang2015improved}
J. Wang
\newblock ``Support recovery with orthogonal matching
pursuit in the presence of noise: A new analysis,''
\newblock{\em IEEE Transactions on Signal processing}, vol. 63, no. 21, pp. 5868--5877,
Jan. 2015.




\bibitem{wu2013exact}
R.~Wu, W.~Huang, and D~Chen,
\newblock ``The exact support recovery of sparse signals with noise via
  orthogonal matching pursuit,''
\newblock {\em IEEE Signal Processing Letters}, vol. 20, no. 4, pp. 403--406,
  Apr. 2013.
  
  
\bibitem{wen2016}
J. Wen, Z. Zhou, J. Wang, X.  Tang and Q. Mo, 
\newblock ``{ A sharp condition for exact support recovery with orthogonal matching pursuit},''
\newblock {\em IEEE Transactions on Signal Processing}, vol. 65, no. 6, pp. 1370--1382, 
  2016.  

\bibitem{livshitz2014sparse}
E.~D. Livshitz and V.~N. Temlyakov,
\newblock ``Sparse approximation and recovery by greedy algorithms,''
\newblock {\em IEEE Trans. Inform. Theory}, vol. 60, no. 7, pp. 3989--4000,
  Jul. 2014.


\end{thebibliography}
\end{document}